\documentclass[12pt]{iopart}
\usepackage[dvips]{graphicx}
\usepackage{cite}
\usepackage{iopams}
\usepackage{subfigure}
\usepackage{wasysym}
\newcommand{\rd}{{\rm d}}
\newcommand{\re}{{\rm e}}
\newcommand{\ri}{{\rm i}}

\newcommand{\abs}[1]{\vert#1\vert}
\newcommand{\be}{\begin{equation}}
\newcommand{\ee}{\end{equation}}
\newcommand{\sech}{{\rm sech}}
\newcommand{\nn}{\nonumber}

\newcommand{\cn}{{\rm cn}}
\newcommand{\sn}{{\rm sn}}
\newcommand{\dn}{{\rm dn}}
\newcommand{\cosech}{{\rm cosech}}

\begin{document}

\title{Bound and resonance states of the nonlinear Schr\"odinger
equation in simple model systems}

\author{D Witthaut, S Mossmann, and H J Korsch}

\address{FB Physik, Technische Universit\"at Kaiserslautern,
D--67653 Kaiserslautern, Germany}

\ead{korsch@physik.uni-kl.de}

\begin{abstract}

\noindent
The stationary nonlinear Schr\"odinger equation,
or Gross--Pitaevskii equation, is studied for the cases of
a single delta potential and a delta--shell potential.
These model systems allow analytical solutions, and thus
provide useful insight into the features of stationary bound,
scattering and resonance states of the nonlinear Schr\"odinger equation.
For the single delta potential, the influence of the potential
strength and the nonlinearity is studied
as well as the transition from bound to scattering states.
Furthermore, the properties of resonance states for a repulsive
delta--shell potential are discussed.
\end{abstract}

\submitto{\JPA}

\maketitle

\section{Introduction}

In the case of low temperatures,
the dynamics of a Bose--Einstein condensate can be described
in a mean--field approach by the
nonlinear Schr\"odinger equation or Gross--Pitaevskii equation
\cite{Lifs80}.
We will focus on the one--dimensional case, which can be achieved
experimentally by a tight confinement in the two other spatial directions
(see for example \cite{Grei01} and references therein).
The nonlinear Schr\"odinger equation for the macroscopic
wavefunction is then given by
\begin{equation}
  \left( -\frac{\hbar^2}{2m} \frac{\partial^2}{\partial x^2} + V(x) +
  g \abs{\psi(x,t)}^2 \right) \psi(x,t) = \ri \hbar \,\frac{\partial \psi(x,t)}{\partial t},
\end{equation}
where $g = 4 \pi \hbar^2 a N /m$ is the nonlinear ``interaction strength'' and
$N$ is the number of particles in the condensate. The wavefunction is
normalized to $\| \psi \| = 1$.
In this ansatz, one only takes elastic $s$--wave scattering into account,
characterized by the $s$--wave scattering length $a$.
The scattering length $a$ and thus the nonlinearity $g$ are negative
for an attractive
nonlinear interaction and positive for a repulsive one.
Another important application of the nonlinear Schr\"odinger
equation is the propagation of electromagnetic waves in
nonlinear media (see, e.g., \cite{Dodd82}, Ch. 8).

Analytic solutions of the nonlinear
equation for a non--vanishing potential $V(x)$ are rare and therefore
it is of interest to study such a simple case in some detail.
Here we study the nonlinear Schr\"odinger equation for two
simple potentials: a single delta potential
\begin{equation}
  V(x) = \lambda \delta(x),
  \label{eqn_delta-potential}
\end{equation}
modelling a short range interaction,
and the delta--shell potential
\be
  V(x) = \left\{
  \begin{array}{*{2}{l}}
  + \infty & \mbox{for} \, x < 0 \\
  \lambda \delta(x-a)   &\mbox{for} \, x \ge 0 \\
  \end{array} \right.
  \label{eqn_dshell-potential}
\ee
with $a > 0$. The delta--shell is a popular model system for
the study of resonances and decay.
We confine ourselves to the stationary case, where the time dependence
is given by the factor $\re^{- \ri \mu t / \hbar}$.
Using units with $\hbar = 1$ and $m=1$,
the stationary nonlinear Schr\"odinger equation reads
\begin{equation}
  \left( -\frac{1}{2} \frac{\rd^2}{\rd x^2} + V(x) +
  g \abs{\psi(x)}^2 \right) \psi(x) = \mu \, \psi(x).
  \label{eqn-GPE}
\end{equation}
The solutions of equation (\ref{eqn-GPE}) for the delta potential
and the delta--shell potential are essentially the ones
of the free nonlinear Schr\"odinger equation. The wavefunction
itself is continuous, but due to the delta potential, its first
derivative is discontinuous at $x = 0$, resp.~$x  = a$:
\be
   \lim_{\epsilon \to 0+}\big( \psi'(a+\epsilon) - \psi'(a-\epsilon)\big)
   = 2 \lambda \psi(a).
  \label{eqn-delta-condition}
\ee
One can easily show that this behaviour, well--known for the
Schr\"odinger equation, is not changed by the nonlinearity.
Furthermore, in the case of the delta--shell potential,
the boundary condition $\psi(0) = 0$ has to be
obeyed.

\section{Single delta potential}
\label{sec-single_delta}

The single delta potential (\ref{eqn_delta-potential}) is
the easiest model for the study of the existence and the
properties of bound and scattering states.
It has been studied in the context of a
nonlinear flow \cite{Haki97,Lebo01}, however rather briefly.

In the linear case, $g=0$, equation (\ref{eqn-GPE}) with $\lambda < 0$
supports a single bound state with energy $E_0 = -\lambda^2/2$ and a
continuous spectrum for $E > 0$, however, without embedded resonances.
The normalized bound state wavefunction is
\be
  \psi_0(x) = \sqrt{|\lambda|} \, \re^{\, \lambda |x|}\,.
  \label{free_sol}
\ee
In the following, we will study the modifications of this linear case due
to an attractive resp.~repulsive nonlinearity.
By means of the scaling $x= x'/|g|$, $\psi=\psi' \sqrt{|g|}$,
$\lambda=\lambda'|g| $ and $\mu=\mu'g^2$ (which conserves the normalization),
the parameter $g$ in (\ref{eqn-GPE}) can be removed up to a sign. Therefore
we will fix the nonlinearity to $g = \pm 1$
(with the exception of section \ref{sec-nonlinearity}).

\subsection{Attractive nonlinearity}
\label{sec-attractive-nonlin}

In the case of an attractive nonlinearity, $g = -1$, the
nonlinear Schr\"odinger equation (\ref{eqn-GPE}) has the well--known bright
soliton solution  for $\lambda =0$ and  $\mu <0$ \cite{Carr00b,Dago00},
\be
  \psi(x) = k \, \sech \big( k\,(x-x_0)\big) \quad \mbox{with}
  \quad k=\sqrt{-2\mu}\,.
 \label{eqn-bright_soliton}
\ee
In order to find nonlinear bound states, i.e.~normalizable solutions
of equation (\ref{eqn-GPE}), bright soliton solutions of the form
(\ref{eqn-bright_soliton}) for $x > 0$ and
$x < 0$ are matched at $x = 0$ by means of condition
(\ref{eqn-delta-condition}).
Obviously, the wavefunction $\psi(x)$ has to be symmetric with respect
to $x=0$ and is therefore given by expression
(\ref{eqn-bright_soliton}) for $x \ge 0$ and
$\psi(x) = \psi(-x)$ otherwise.
Inserting this ansatz into equation (\ref{eqn-delta-condition}) leads
to the condition
\be
   \tanh ( k x_0)
   = \lambda/k\,. \label{eqn-mux0}
\ee
Combined with the normalization of the wavefunction,
\begin{eqnarray}
1&=&\int_{-\infty}^{+\infty}\big|\psi(x)\big|^2\,\rd x
=2k^2\int_0^{+\infty}\sech^2 \big( k\,(x-x_0)\big)\,\rd x \nonumber\\[2mm]
&=&2k\big(1+\tanh(kx_0)\big)\,,
\end{eqnarray}
this yields
\be
   k = {\textstyle \frac{1}{2}} - \lambda\ ,\quad \mbox{i.e.}\quad
   \mu=-{\textstyle \frac{1}{8}} (2\lambda-1)^2\,.
   \label{eqn-brightsol_mu}
\ee
Because of $|\tanh(k x_0)| < 1$, one finds a condition
for the existence of a bound state:
\be
  \lambda < \lambda_c = {\textstyle \frac{1}{4}} \,.
\ee
A bound state exists for any attractive
delta potential but also for a repulsive one, provided that
its strength is not too large. This effect is due to the attractive
self interaction $-|\psi(x)|^2$ which can compensate a limited
external repulsion.

\begin{figure}[ht]
\begin{center}
\includegraphics[width=7cm,  angle=0]{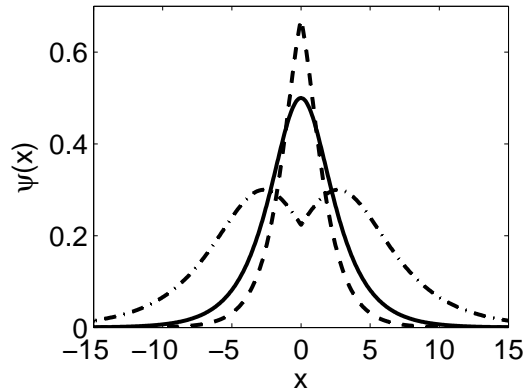}
\end{center}
\caption{\label{fig-bright_soliton1}
Wavefunctions $\psi(x)$ of bound states of the nonlinear
Schr\"odinger equation (attractive nonlinearity, $g=-1$)
for a delta potential $V(x) = \lambda \delta(x)$
with  $\lambda = -0.2$ ($- - -$), $\lambda = 0$ (---)
and $\lambda = +0.2$ ($- \cdot - $).}
\end{figure}
Figure \ref{fig-bright_soliton1} shows the wavefunctions
for such bound states for three different values of the
potential strength $\lambda$.
Quite generally, for an attractive delta potential
with a negative value of $x_0$,
the wavefunction tends to concentrate at the delta potential
with decreasing $\lambda$. A repulsive delta potential
repels the wavefunction, $x_0$ is positive, and one observes two peaks
of $\psi(x)$ at $x= \pm x_0$
that are pushed further away as $\lambda$ is increased toward
$\lambda_c$.
For $\lambda \rightarrow \lambda_c$
the wavefunction evolves into two infinitely separated bright
soliton solutions.

Remarkably, the bound state ceases to exist at a finite negative value
of the chemical potential
\be
   \mu_c=-{\textstyle \frac{1}{8}} (2\lambda_c-1)^2 =
   -{\textstyle \frac{1}{32}} \,.
\ee
This difference to the linear equation or the case of repulsive
nonlinearity (see below) corresponds to the fact that the
wavefunction is no longer bound by an external potential but
by the internal self--interaction.

For $\lambda > \lambda_c$, there is no bound state solution any more,
but one can actually find periodic stationary solutions in terms of Jacobi
elliptic functions \cite{Carr00b,Dago00}
\be
  \psi(x) = \frac{\sqrt{p}ß,4 K(p)}{L} \, \cn \bigg(4 K(p)\frac{x-x_0}{L}
  \,\bigg|\, p \bigg).
  \label{eqn-bright_periodic_solution}
\ee
Here $L$ is the period, $p \in [0,1]$ the elliptic modulus
of the Jacobi elliptic function $\cn $ and $K(p)$ denotes the complete
elliptic integral of the first kind. The chemical potential is
related to these parameters by
\be
  \mu = 8(1-2p) \, K^2(p)/L^2\,.
\ee
These solutions are of course no longer normalizable, and will be denoted
as scattering states in the following.
Such a periodic solution, characterized by three parameters, the chemical
potential $\mu$, the period $L$ and the shift $x_0$, has to
fulfil only condition (\ref{eqn-delta-condition}).
Thus, for a fixed value of the potential strength $\lambda$, there
exists a variety of solutions, whereby the chemical potential $\mu$
and the period $L$ can be chosen more or less independently.
The value of $x_0$ is then fixed to satisfy condition
(\ref{eqn-delta-condition}).

\begin{figure}[ht]
\begin{center}
\includegraphics[width=7cm,  angle=0]{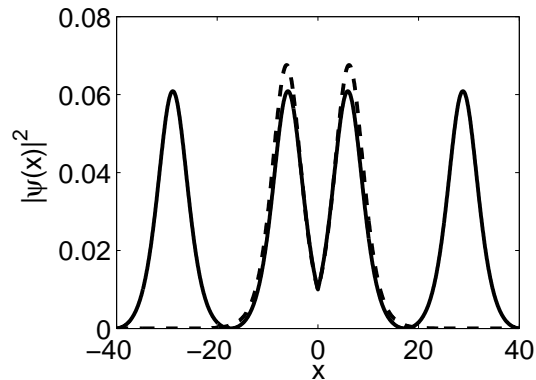}
\end{center}
\caption{\label{fig-dpot-transition1_wavefun}
Wavefunctions of a bound state ($- -$, $\lambda = 0.24$)
and a scattering state (---, $\lambda = 0.26$) for a repulsive
delta potential $V(x) = \lambda \delta(x)$
for an attractive nonlinearity close to the transition at $\lambda_c=0.25$.}
\end{figure}

In the following, we discuss a particular class of solutions that merge
continuously into the bound state solution when $\lambda$ is
decreased below its critical value $\lambda_c$.
Therefore we make the ansatz that $\mu$ and
$\psi(x = 0)$ depend continuously on the strength $\lambda$ of the
delta potential at $\lambda_c$. In fact, we assume the functional
relation to be the same for $\lambda > \lambda_c$ and
$\lambda < \lambda_c$, i.e.~given by equation (\ref{eqn-brightsol_mu}),
and
\be
  \psi(0) = k \, \sech \left( {\rm arctanh}
  (\lambda/k) \right),
  \label{eqn-bs-psi0-fixed}
\ee
respectively. For a given value of $\lambda$, we construct solutions
(\ref{eqn-bright_periodic_solution}) that fulfil condition
(\ref{eqn-delta-condition}) and yield the desired values of
$\mu$ and $\psi(0)$. Such solutions can indeed be found
and figure \ref{fig-dpot-transition1_wavefun} illustrates
such a wavefunction for $\lambda = 0.26$, just above
the critical value $\lambda_c=0.25$, in comparison to a bound state solution
for $\lambda = 0.24$.

In the vicinity of the delta potential at $x = 0$, both wavefunctions
look rather similar and thus
the transition from a bound to a scattering state seems to be
continuous. The observed difference between the bound state and the
periodic solution for $|x|>L$ disappears in the limit
$\lambda \searrow \lambda_c$ because the period $L$ of the
Jacobi elliptic solution moves toward infinity.

\begin{figure}[htb]
\begin{center}
\includegraphics[width=6.5cm,  angle=0]{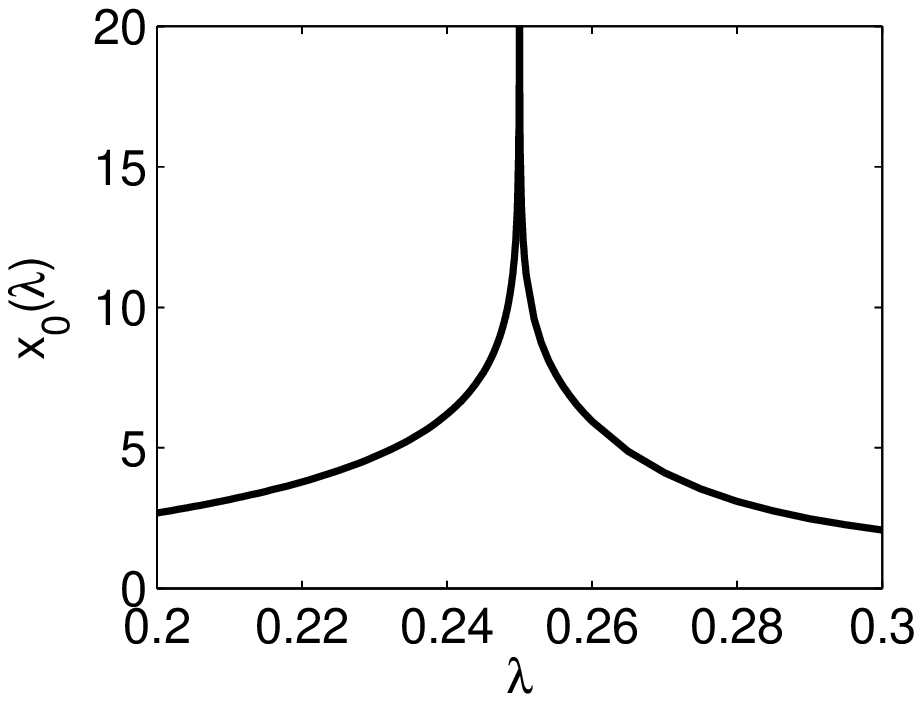}
\hspace{2mm}
\includegraphics[width=6.5cm,  angle=0]{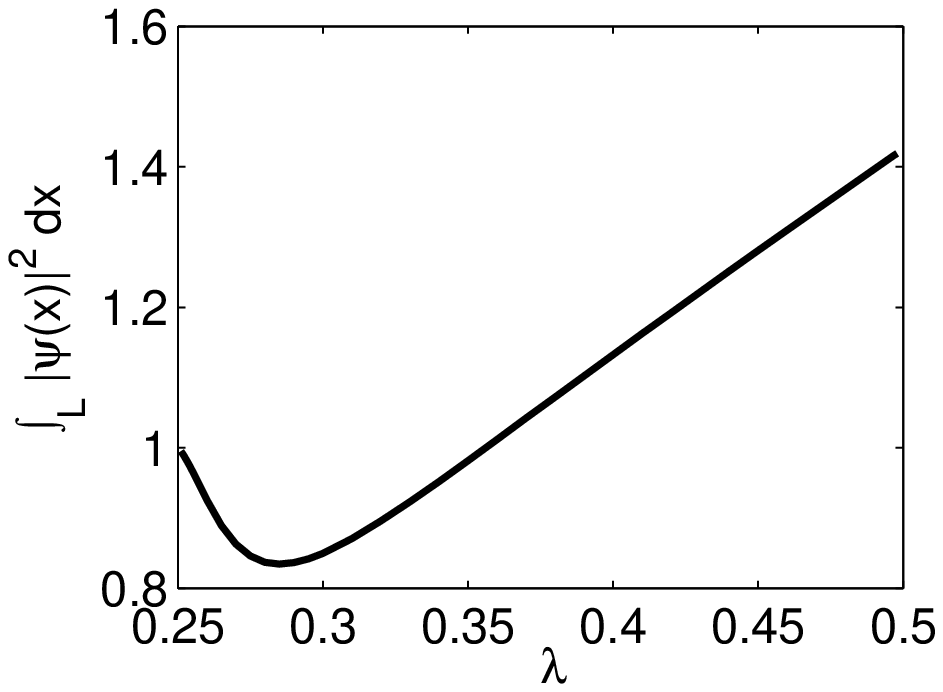}
\end{center}
\caption{\label{fig-dpot-transition1_max}
Transition from a bound to a scattering state. Left:
Position $x_0(\lambda)$ of the first maximum of the wavefunction.
Right: Norm per period $L$. Note that $L \to \infty$
as $\lambda \searrow \lambda_c$.}
\end{figure}

To explore this transition in some detail, we consider
the position $x_0$ of the first maximum of $|\psi(x)|^2$,
as a function of $\lambda$, given by
\be
  x_0(\lambda) = \frac{1}{1/2- \lambda} \, {\rm arctanh}
  \left( \frac{\lambda}{1/2-\lambda} \right)
\ee
for $\lambda < \lambda_c$ and by the solution of the complex equations
\begin{eqnarray}
  \psi(0) &=& k \sqrt{\frac{p}{2p-1}} \; \cn \bigg(\frac{k x_0(\lambda)}{
  \sqrt{2p-1}}   \,\bigg|\, p \bigg) \quad {\rm and} \nonumber \\
  \lambda \psi(0) &=& k^2 \frac{\sqrt{p}}{2p-1} \;  \sn \bigg(\frac{k
  x_0(\lambda)}{ \sqrt{2p-1}}\,\bigg|\, p \bigg) \; \dn \bigg(\frac{k
  x_0(\lambda)}{ \sqrt{2p-1}}\,\bigg|\, p \bigg)
\end{eqnarray}
for $\lambda > \lambda_c$,
where $k$ and $\psi(0)$ are fixed by equations
(\ref{eqn-brightsol_mu}) and (\ref{eqn-bs-psi0-fixed})
as discussed above.
At $\lambda_c$, the function
$x_0(\lambda)$ shown in figure \ref{fig-dpot-transition1_max}
on the left has a logarithmic singularity.

On the right of figure \ref{fig-dpot-transition1_max},
the norm per period $\int_{-L/2}^{L/2} |\psi(x)|^2 \rd x$
is displayed, which tends to unity at the critical
point $\lambda_c$, i.e.~it approaches the
bound state normalization. Hence the norm is also continuous.

\subsection{Repulsive nonlinearity}
\label{sec-repulsive-nonlin}
In the case of a repulsive nonlinearity, $g = +1$, the
nonlinear Schr\"odinger equation has the well--known dark
soliton solutions for $\lambda =0$ \cite{Carr00a,Dago00}:
\be
  \psi = \sqrt{\mu} \, \tanh{\left( \sqrt{\mu}
 (x-x_0)\right)}.
 \label{eqn-dark_soliton}
\ee
Making such an ansatz separately for $x > 0$ and $x < 0$ and matching
at $x = 0$ with respect to condition
(\ref{eqn-delta-condition}) yields $x_0 = 0$ regardless of the
value of $\lambda$.
Remarkably, the wavefunction has a zero at $x = 0$ even for an
attractive delta potential.
But solutions of this kind are of course not normalizable.
Another possible solution is
\be
  \psi(x) = k \, \cosech ( k\, (x-x_0)) \ ,\quad \mbox{with}
  \quad k=\sqrt{-2\mu}\,,
 \label{eqn-cosech_soliton}
\ee
which is usually discarded because of its unphysical singularity at $x = x_0$.
In the case of a delta potential, however, this ansatz reveals
proper stationary bound states.
Assuming (\ref{eqn-cosech_soliton}) for $x > 0$, a short calculation shows
that the wavefunction has to be symmetric, $\psi(-x) = \psi(x)$.
In addition, $x_0$ must be negative because otherwise the wavefunction would
become singular at $x=x_0$.
Condition (\ref{eqn-delta-condition}) yields
\be
   \tanh (k x_0)
   = k/\lambda \label{eqn-mux1}
\ee
and the normalization of the wavefunction requires
\begin{eqnarray}
1&=&\int_{-\infty}^{+\infty}\big|\psi(x)\big|^2\,\rd x
=2k^2\int_0^{+\infty}\!\!\cosech^2 \big( k\,(x-x_0)\big)\,\rd x\nonumber\\[2mm]
&=&-2k\big(1+\coth(kx_0)\big)\,.
\end{eqnarray}
This leads to
\be
  k = -{\textstyle \frac{1}{2}}  - \lambda
\ee
which must be positive, yielding the condition
\be
  \lambda < \lambda_c = -{\textstyle \frac{1}{2}}\,,
\ee
i.e.~the delta potential must be sufficiently attractive to
overcome the repulsive self--interaction in order to support
a bound state.

\begin{figure}[htb]
\begin{center}
\includegraphics[width=7cm,  angle=0]{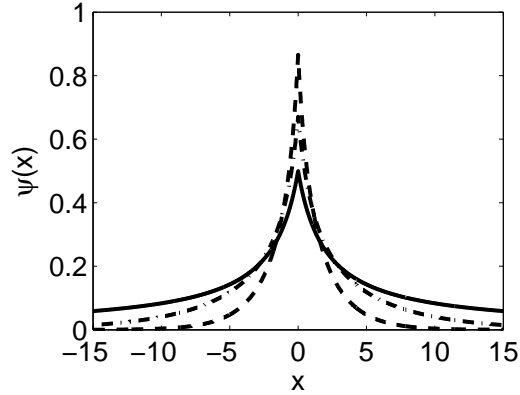}
\end{center}
\caption{\label{fig-rep_soliton1}
Wavefunctions $\psi(x)$ of bound states of the
nonlinear Schr\"odinger equation (repulsive nonlinearity, $g=+1$)
for a delta potential $V(x) = \lambda \delta(x)$ for
three values of the potential strength $\lambda = -1$ ($- - -$),
$\lambda = -0.7$
($-\cdot -$) and for the critical case $\lambda=-0.5$ (---)\,.}
\end{figure}
In figure \ref{fig-rep_soliton1}, such bound states are displayed
for different values of the potential strength $\lambda$.
For decreasing values of $\lambda$, the wavefunction concentrates
at the position of the  delta potential.
In the opposite limit, $\lambda \nearrow \lambda_c$, we see by
series expansion of the $\tanh$ and the $\sinh$ functions, that
\be
  x_0 \to x_c=1/\lambda_c \quad \mbox{and}\quad
  \mu \to \mu_c=0
\ee
and that the wavefunction converges to the limiting function
\be
  \psi_c(x)=\frac{1}{|x|-x_c}=\frac{1}{|x|+2}
  \label{eqn-bs-rep-critical}
\ee
also shown in figure \ref{fig-rep_soliton1}.
This is in contrast to the case of an attractive nonlinearity where the
bright soliton peaks move to $\pm \infty$ at the critical
value $\lambda_c$.

For $\lambda > \lambda_c$, one again finds periodic solutions in terms
of Jacobi elliptic functions \cite{Carr00a,Dago00}
\be
  \psi(x) = \frac{\sqrt{p}\,4 K(p)}{L} \, \sn \bigg(4 K(p)\frac{x-x_0}{L}
  \,\bigg|\, p \bigg),
  \label{eqn-dark_periodic_solution}
\ee
where $L$ is the periodicity, $p \in [0,1]$ the elliptic modulus
of the Jacobi elliptic function and $K(p)$ denotes the complete
elliptic integral of the first kind.
The  chemical potential is given by
\be
  \mu = 8(p+1) \, K(p)^2/L^2.
  \label{eqn-rep-periodic-mu}
\ee
For a fixed value of the potential strength $\lambda$, one again finds
a variety of solutions, whereby the chemical potential $\mu$
and the period $L$ can be chosen more or less independently.
Note that such periodic solutions can only be found
for $\mu > 0$.

Nevertheless, one can find a lower bound for the period $L$. From
equation (\ref{eqn-rep-periodic-mu}) it follows that
\be
  L  = \sqrt{\frac{8(p+1)}{\mu}} \, K(p) \ge \frac{2 \pi}{\sqrt{2 \mu}} \, .
\ee
For $\lambda \searrow \lambda_c$ and $\mu \searrow 0$ the period of
the wavefunction $L$ tends to infinity and the wavefunction
is not periodic any more in this limit.
But in this case one cannot find a continuous transition to the bound
state wavefunction (\ref{eqn-cosech_soliton}).
For $\lambda \nearrow \lambda_c$ one finds the bound state
(\ref{eqn-bs-rep-critical}) with $\psi(0) = 1/2$ and $\mu = 0$.
In contrast, we have $\psi(0) \rightarrow 0$ for $\mu \searrow 0$
for the periodic solution (\ref{eqn-dark_periodic_solution})
because of equation (\ref{eqn-rep-periodic-mu}).
In fact, the elliptic function $\sn$ evolves continuously
into the $\tanh$ when the elliptic modulus $p$ tends to unity
\cite{Abra72}.

Actually, there exist Jacobi elliptic functions that merge
continuously into the cosech as the elliptic modulus $p$
tends to one. These solutions are given in terms
of the Jacobi elliptic functions ds and cs \cite{Abra72}.
But these functions have poles at the zeros of the $\sn$
and thus are not physical.
\subsection{Variation of the nonlinearity}
\label{sec-nonlinearity}

In this section, we will briefly discuss the influence of the mean--field
interaction strength, i.e.~the nonlinearity $g$, on the solutions of the
nonlinear Schr\"odinger equation for an attractive delta potential. We
therefore reintroduce the parameter $g$.

\begin{figure}[h]
\begin{center}
\includegraphics[width=7cm,  angle=0]{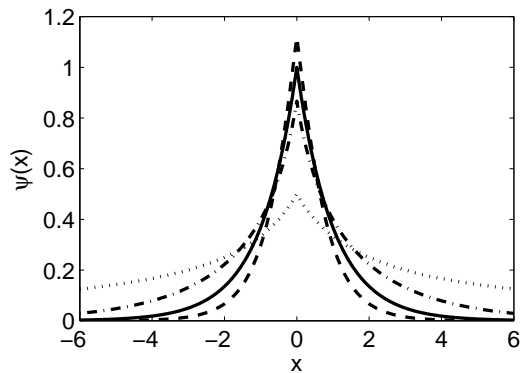}
\end{center}
\caption{\label{fig-wavefun-nonlin}
Bound state wavefunctions $\psi(x)$ of the nonlinear
Schr\"odinger equation for an attractive delta potential
$V(x) = -\delta(x)$ for different values of the nonlinearity:
$g = -1$ ($- -$), $g=0$ (---), $g=+1$ ($-\cdot -$) and  $g=g_c=2$
($\cdot \cdot \cdot$)\,.}
\end{figure}
The bound state solutions have already been deduced in the previous
sections.
In figure \ref{fig-wavefun-nonlin}, the wavefunction of such a bound
state is displayed for three different values of the nonlinearity
$g = -1, \, 0, \, +1$ and a fixed potential strength $\lambda = -1$.
With increasing nonlinearity $g$, the wavefunction is pushed
outward.

In both cases of attractive and repulsive nonlinearity, the
chemical potential is given by
\be
  \mu = - \frac{1}{8} \left( 2 \lambda + g \right)^2 \, ,
\ee
which follows directly from the matching condition
(\ref{eqn-delta-condition}) and the normalization of the wavefunction.
At a critical value of $g$, the chemical potential $\mu$ becomes
zero and the bound state ceases to exist.
The condition for the existence of a bound state is the same
as discussed in section \ref{sec-repulsive-nonlin}. Reformulated in
terms of the nonlinearity parameter $g$, it reads:
\be
  g < g_c = -2 \lambda\,.
\ee
When $g$ approaches the critical value $g_c$, the situation is
similar to the case of a fixed repulsive nonlinearity $g$ and
$\lambda \nearrow \lambda_c$ as discussed in the previous section.
The wavefunction at the critical value of $g$ is
\be
  \psi_c(x)=\frac{\sqrt{-\lambda}}{|x|-2\lambda}\,.
\ee

\begin{figure}[tb]
\begin{center}
\includegraphics[width=6cm,  angle=0]{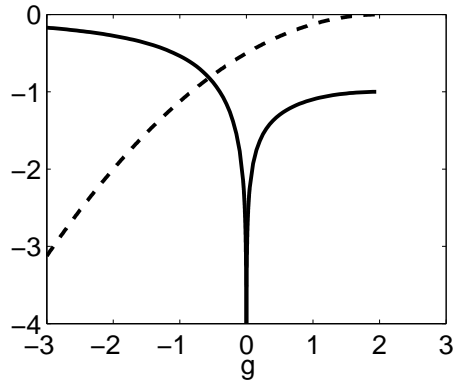}
\end{center}
\caption{\label{fig-mu-width-nonlin}
Dependence of the chemical potential $\mu$ ($- -$) and the parameter
$ x_0$ (---) on the nonlinearity $g \le g_c = 2$
for the bound state of the nonlinear Schr\"odinger
equation for an attractive delta potential
$V(x) = -\delta(x)$.}
\end{figure}

The dependence of $x_0$ and $\mu$ on the nonlinearity $g$
is illustrated in figure \ref{fig-mu-width-nonlin}.
The position $x_0$ is given by equation (\ref{eqn-mux0})
for $g < 0$ and (\ref{eqn-mux1}) for $g > 0$, however with
$k = - \lambda - g/2$.

For $g=0$, one finds the well--known value $\mu = -\lambda^2/2$,
whereas the function $x_0(g)$ has a logarithmic singularity.
Nevertheless, the bound state wavefunction $\psi(x)$ evolves smoothly
into the well--known bound state (\ref{free_sol}) of the linear
problem for an attractive as well as a repulsive nonlinearity.

For $g \rightarrow g_c = -2 \lambda$, the chemical potential $\mu$
tends to zero and $x_0$ tends to the finite
value $1/\lambda$.
The disappearance of the bound state if $g$ is increased above $g_c$
is similar to the effect observed by Moiseyev {\it et al.}
\cite{Mois03} for a smooth potential $V(x)$ where a bound state is
transformed into a resonance--like state at a critical nonlinear
interaction.

\section{Delta--shell potential}

In this section we discuss another simple and very popular
model system: the delta--shell potential.
A detailed discussion of the linear three--dimensional
delta--shell potential can be found in \cite{Gott66}. Here
we restrict ourselves to the one--dimensional case
\be
  V(x) = \left\{
  \begin{array}{*{2}{l}}
  + \infty & \mbox{for} \, x < 0 \\
  \lambda \delta(x-a)   &\mbox{for} \, x \ge 0 \\
  \end{array} \right.
  \label{eqn_delta-shell-potential}
\ee
with $a > 0$.
First we briefly resume the basic features of the delta--shell potential
in the linear case ($g = 0$), in particular the existence of bound states
in an attractive potential and resonances in a repulsive one.
As we have already discussed the properties of bound states in a single
delta potential in some detail, we now concentrate on the case of
a repulsive potential ($\lambda > 0$).
We set $\hbar=1$ and $m = 1$ as above.

\subsection{The linear case}

In the linear case $g=0$, the wavefunction in a delta--shell potential
is given by
\be
  \psi_k(x) = \left\{\begin{array}{*{2}{l}}
     \sin(kx) & \mbox{for} \; x < a \\ [2mm]
     \sin(kx) + \frac{2\lambda}{k} \sin(ka) \sin(k(x-a))
        & \mbox{for} \; x > a, \\
     \end{array}\right.
\ee
The phase shift $\delta(k)$ between incoming and outgoing
waves for $x>a$ is easily calculated and yields
\be
  \tan \delta(k) = \frac{\cos(2ka) - 1}{\sin(2ka) + k/\lambda} \, .
  \label{eqn-dpot_lin_phase}
\ee
The S--matrix $S(k)$ is defined in terms of the the phase shift $\delta(k)$
by \cite{Tayl72}:
\be
  S(k) = \frac{1+\ri \tan \delta(k)}{1- \ri \tan \delta(k)} \, .
  \label{eqn-smatrix_phase}
\ee

Bound states correspond to poles of the S--Matrix $S(k)$ on the
positive imaginary axis.
Calculating these poles one arrives at
\be
  \re^{2 \ri k a} = 1 - \frac{\ri k}{\lambda} \, .
  \label{eqn-dshell_lin_condition_res}
\ee
This equation has a solution on the positive imaginary axis if
the condition
\be
  \lambda a > - \, \frac{1}{2}
  \label{eqn-dshell_lin_conda}
\ee
is fulfilled. This implies that the delta--shell potential has to
be sufficiently attractive to support a bound state.
If the distance $a$ is reduced or $\lambda$ is increased, so that the condition
(\ref{eqn-dshell_lin_conda}) is not fulfilled any longer, the bound
state is lost and one finds a virtual state instead.
A virtual state corresponds to a pole of the S--matrix $S(k)$ on the
negative imaginary axis \cite{Tayl72}. The wavefunction of such
a state diverges exponentially.
For $a \rightarrow \infty$ the delta--shell potential is equivalent
to a single delta potential and the energy is
$E \rightarrow -\lambda^2/2$.

Naturally there exist no bound states in a repulsive delta--shell potential,
but one can find resonance states.
A resonance is defined  by a pole of the S--matrix $S(k)$ in the lower
half plane \cite{Tayl72}. The energy of the n--th resonance is also complex
\be
  {\cal E}_n = k_n^2/2 = E_n - \ri \Gamma_n/2,
\ee
where the imaginary part $\Gamma_n$ is interpreted as a decay rate.
In the vicinity of a resonance, the phase shift $\delta(k)$ rapidly
changes by an amount of $\pi$.

\begin{figure}[htb]
\centering
\includegraphics[width=7cm,  angle=0]{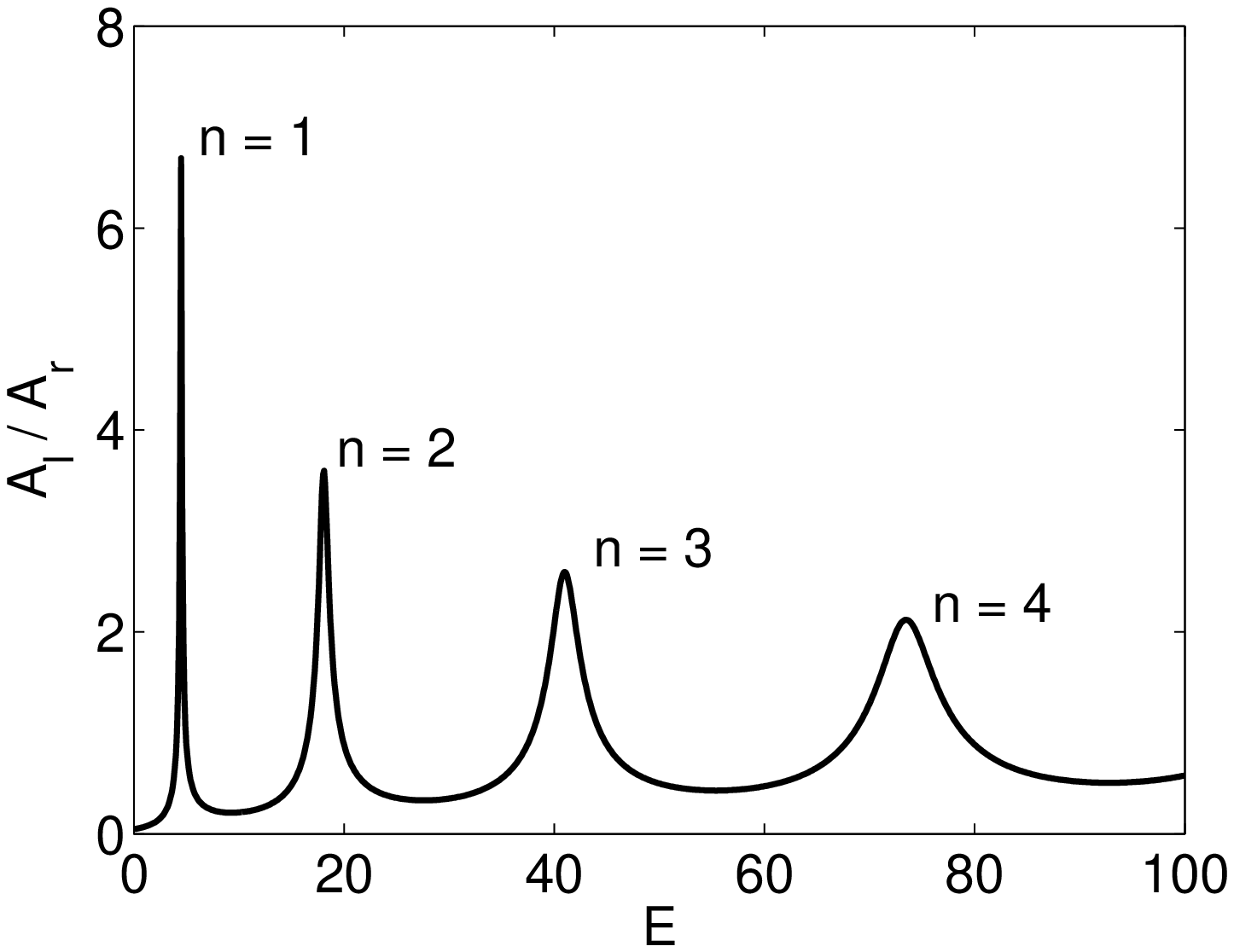}
\hspace{1mm}
\includegraphics[width=7cm,  angle=0]{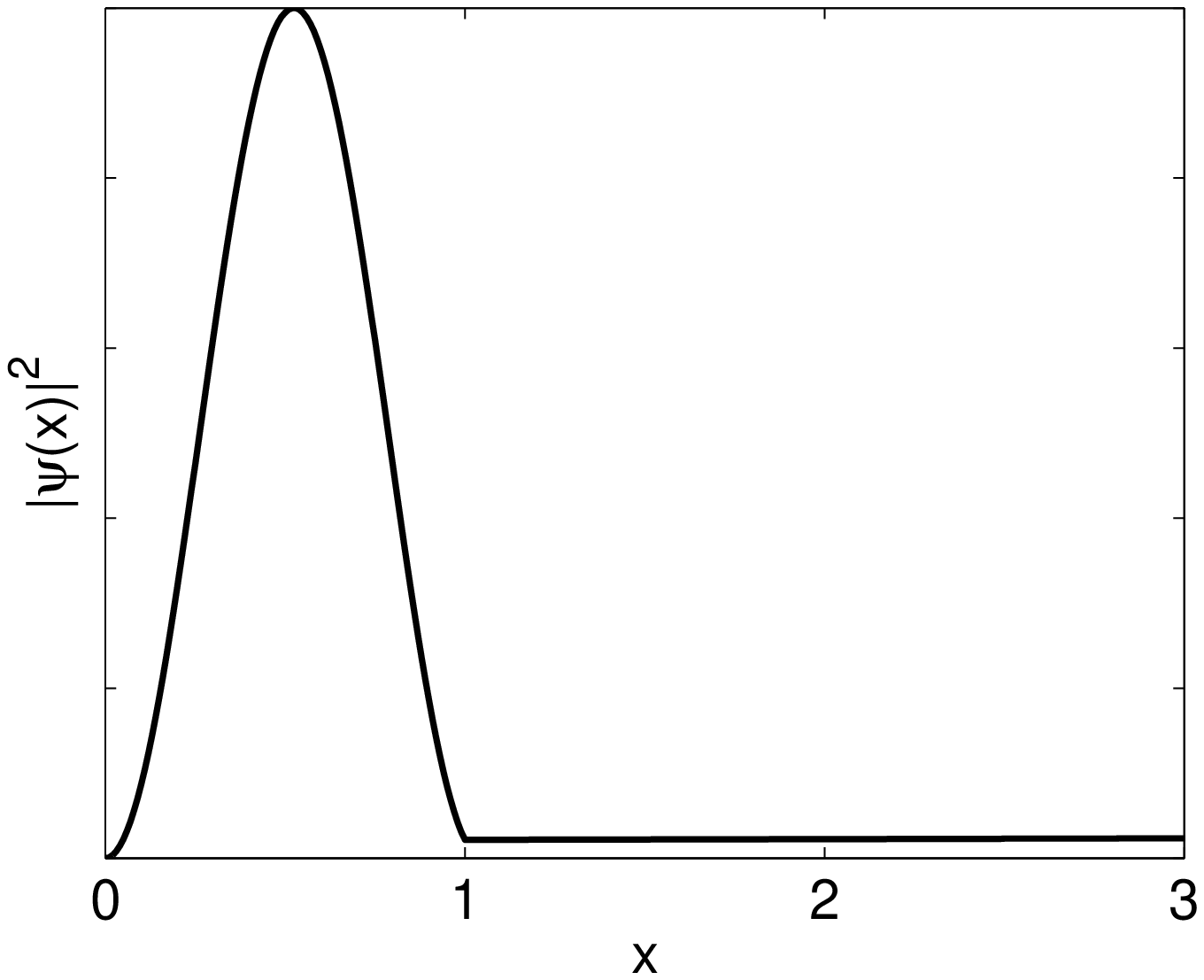}
\caption{\label{fig_dshell_lin_resonance}
Left: Amplitude ratio $A_l/A_r$ for a delta--shell potential
with $a=1$ and $\lambda = 10$ in the linear case.
Right: Squared modulus of the wavefunction of the most stable
resonance at ${\cal E}_1 = 4.488 - 0.063 \ri$.}
\end{figure}

The amplitude of a resonance wavefunction is enhanced for $x < a$.
This is illustrated in figure \ref{fig_dshell_lin_resonance}
for a delta--shell potential of strength $\lambda = 10$ at $a=1$.
The ratio of the amplitudes on the left--hand side ($x < a$) and on the 
right--hand side ($x > a$) of the delta--shell potential, denoted as 
$A_l$ resp.~$A_r$, is plotted for real values of the energy.
The peaks of the amplitude ratio $A_l/A_r$ close to the resonances
are clearly visible.
The squared modulus of the wavefunction of the most stable resonance at
${\cal E}_1 = 4.488 - 0.063 \ri$ is displayed on the right.
Nevertheless one has to keep in mind that the wavefunction finally
diverges exponentially for complex energies ${\cal E}$, whereas
it is periodic for real energies.

\subsection{Resonances in the nonlinear case}

Now we come back to the nonlinear Schr\"odinger equation
\be
  \left( -\frac{1}{2} \frac{\rd^2}{\rd x^2} + \lambda \delta(x-a) +
  g \abs{\psi(x)}^2 \right) \psi(x) = \mu \, \psi(x) \quad \mbox{for} \quad x \ge 0.
\ee
In the following we will only discuss the case of a repulsive
delta--shell potential ($\lambda > 0$).
By means of a scaling $x= x'/s$, $\psi=\psi' \sqrt{s}$,
$\lambda=\lambda's $ and $\mu=\mu' s^2$ for $s>0$
(which conserves the normalization), the number of independent
parameters is reduced to two.
As we are mainly interested in the effects of a varying nonlinearity,
the potential is fixed by $a=1$ and $\lambda = 10$ in the following
examples.
In the linear case we find resonances for this potential.
Now we want to identify and characterize resonances in the
nonlinear case as well.

But the definition of a resonance becomes somewhat ambitious in the
nonlinear case. A decomposition into incoming and outgoing waves
and thus a definition of the S--matrix $S(k)$ is not possible.
One method widely used to compute resonances in the linear case 
is exterior complex scaling (see e.g. \cite{Mois98}). This technique
has also been successfully applied to the nonlinear
Schr\"odinger equation \cite{Mois03,Schl04}.

We will not adopt this approach but rather look for
solutions that can be expressed analytically.
We have already learned that the real solutions of the free
nonlinear Schr\"odinger equation are given in terms of Jacobi
elliptic functions. These solutions are matched at $x=a$ to
obtain solutions for the delta--shell potential.
The chemical potential $\mu$ of such a solution is real.
Thus we can define a resonance neither by a complex eigenenergy
nor via the S--matrix. In the following we will rather call a state
a resonance, when its amplitude is resonantly enhanced in the
vicinity of the potential, i.e.~for $x < a$.

Let us briefly discuss the time evolution of nonlinear resonances.
Note that the states
\be
  \psi(x,t) = \exp(-\ri \mu t) \, \psi(x)
  \label{eqn-gpe-stat-state}
\ee
with a complex chemical potential $\mu = \mu_r - \ri \Gamma/2$
do {\it not} fulfill the time--dependent nonlinear Schr\"odinger
equation, because the norm of these states is not constant.
One can circumvent this problem by introducing an additional
source term or one considers the states (\ref{eqn-gpe-stat-state})
just as an adiabatic approximation \cite{Schl04}.
On the contrary the states (\ref{eqn-gpe-stat-state}) with a real
chemical potential $\mu$ discussed in this paper fulfill
the time--dependent nonlinear Schr\"odinger equation but
do {\it not} decay.

Furthermore we have to be cautious about the nonlinear
parameter $g$. A meaningful definition of the nonlinearity
requires that the norm or the amplitude of a solution must
be fixed in some way, e.g.~by $\|\psi\| = 1$ in section
\ref{sec-single_delta}.
This is not applicable any longer since resonance states
are not normalizable.
As a global measure of the nonlinear interaction we thus
define the mean--field potential $g |\psi(x)|^2$, integrated
over the "interaction--region" $0<x<a$ of the external potential:
\be
  g_{\rm eff} = g \int_0^a |\psi(x)|^2 \, \rd x \, .
  \label{eqn-geff}
\ee

\subsection{Attractive Nonlinearity}

First we discuss the nonlinear Schr\"odinger equation with
a negative nonlinearity $g$, corresponding to an attractive
mean--field interaction.
As stated above, the real--valued periodic solutions of the free nonlinear
Schr\"odinger equation with a negative nonlinearity can be
expressed in terms of the Jacobi elliptic function cn
\cite{Carr00b, Dago00}.
Thus, in order to find solutions for the delta--shell potential we
make an ansatz of the form (\ref{eqn-dark_periodic_solution})
separately for $x<a$ and $x>a$:
\be
\psi(x) = \left\{ \begin{array}{l}
 \psi_l(x) = A_l \, \cn \Big( 4 K(p_l) \big( \frac{x}{L_l} +\frac{1}{4}
    \big) \,\Big|\, p_l \Big) \quad \mbox{for} \; x < a \\ [2mm]
 \psi_r(x) = A_r \, \cn \Big( 4 K(p_r)  \frac{x+x_0}{L_r}
    \,\Big|\, p_r \Big) \qquad \mbox{for} \; x > a.
  \end{array} \right.
  \label{eqn-dshell-attres-ansatz}
\ee
The amplitudes $A_{l,r}$ and the periods $L_{l,r}$ are given by
\be
  A_{l,r} = \frac{4 \, \sqrt{p_{l,r}} \, K(p_{l,r})}{\sqrt{|g|} \, L_{l,r}} \quad \mbox{and} \quad
  \mu = \frac{8 \, (1 - 2p_{l,r}) \, K(p_{l,r})^2}{L_{l,r}^2} \, ,
  \label{eqn-dshell-attres-amp-mu}
\ee
where $p_{l,r}$ are the elliptic parameters of
the solution on the left--hand ($x<a$) and on the right--hand ($x>a$)
side of the delta--shell.
Clearly one has only one value for the chemical potential,
whereas the amplitude $A$, the parameter $p$ and the period $L$
generally differ for $x <a$ and $x>a$. This is different from
the linear case, where the period $L$ is fixed with the energy.
The chemical potential is positive, $\mu \ge 0$, if the elliptic
parameter is restricted to $p_{l,r} \in [0,1/2]$.

The boundary condition $\psi(0) = 0$ is automatically fulfilled
by this ansatz.
Furthermore the wavefunction must be continuous at $x=a$,
whereas its derivative is discontinuous according to equation
(\ref{eqn-delta-condition}), leading to the conditions:
\begin{eqnarray}
\fl {\rm I.}  \qquad  A_l \, \cn ( u_l | p_l ) &=& A_r \, \cn (u_r  | p_r ) \\[2mm]
\fl  {\rm II.} \quad 2 \lambda A_l \, \cn(u_l  | p_l )
   &=& - \frac{4 A_r K_r}{L_r} \sn (u_r  | p_r ) \dn (u_r  | p_r ) 
   + \frac{4 A_l K_l}{L_l} \sn (u_l  | p_l ) \dn (u_l  | p_l ),
  \label{eqn-dshell-attres-conditions}
\end{eqnarray}
where the abbreviations $u_l = K(p_l) (4 a/L_l +1)$ and
$u_r = 4  K(p_r) (a+x_0)/L_r$ have been used.
The first condition can be fulfilled by an appropriate choice
of $x_0$, as long as $|A_l \, \cn(u_l|p_l)| \le |A_r|$. Then one
can insert the first condition into the second one and arrives at
\begin{eqnarray}
 &&\frac{2 \lambda^2}{\mu} \frac{p_l}{1-2 p_l} \cn(u_l|p_l)^2
   - \frac{4 \lambda}{\sqrt{2\mu}} \frac{p_l}{(1-2p_l)^{3/2}}
     \, \cn(u_l|p_l) \dn(u_l|p_l) \sn(u_l|p_l) \nn \\ [2mm]
     && = \frac{(1-p_r)p_r}{(1-2p_r)^2} - \frac{(1-p_l)p_l}{(1-2p_l)^2}\, .
 \label{eqn-dshell-attres-cond2}
\end{eqnarray}

As argued above we are looking for solutions whose amplitudes
are resonantly enhanced for $x<a$, i.e.~for solutions with
a maximum amplitude ratio $A_l/A_r$.
This ratio is given directly by the elliptic parameters
$p_{l,r}$:
\be
  \frac{A_l}{A_r} = \left[ \frac{p_l (1-2p_r)}{(1-2p_l)p_r} \right]^{1/2}.
  \label{eqn-dshell-attres-ampratio}
\ee
A resonant enhancement of the amplitude ratio demands that $p_l \gg p_r$.

\begin{figure}[htb]
\centering
\includegraphics[width=7cm,  angle=0]{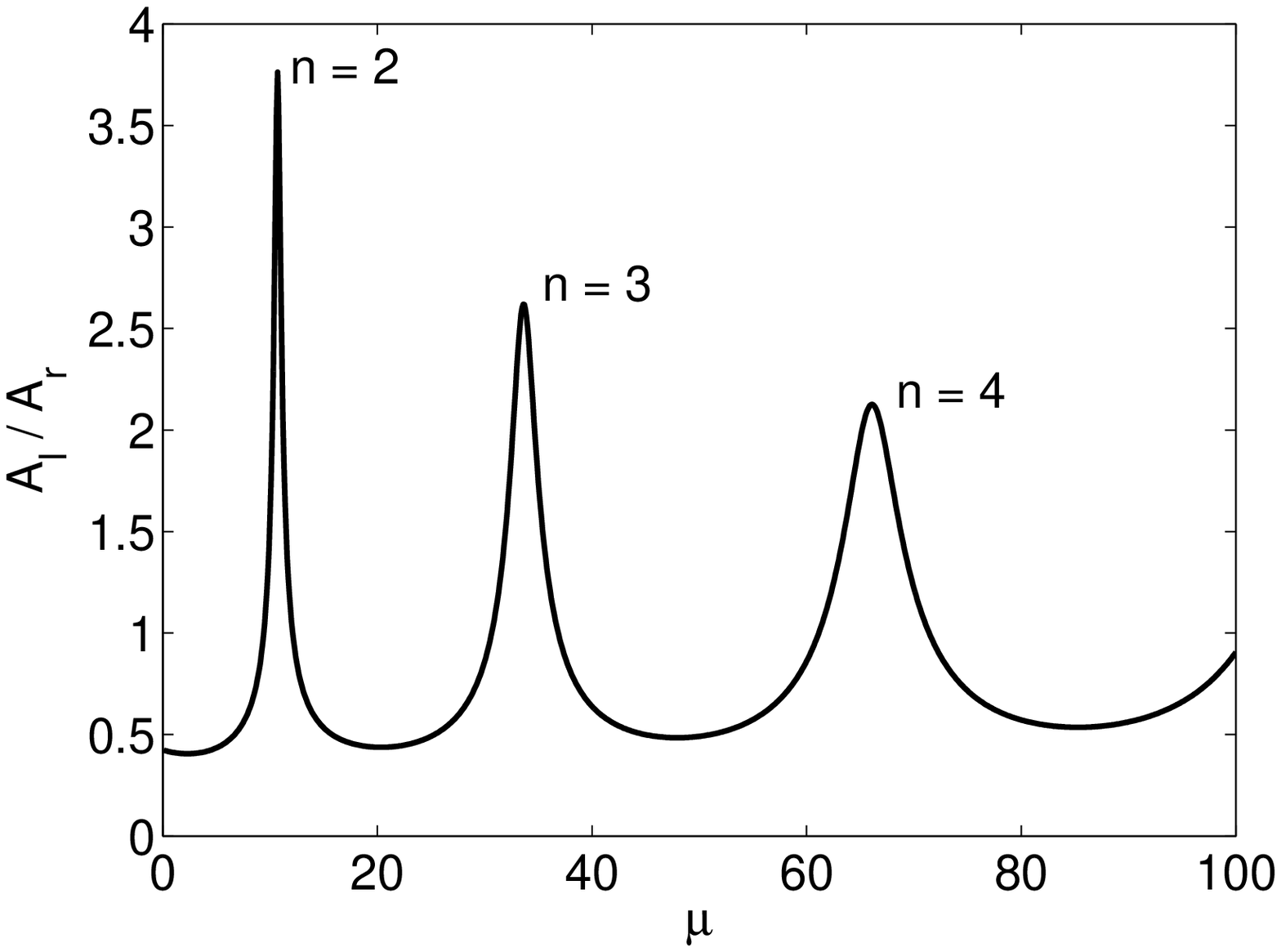}
\hspace{5mm}
\includegraphics[width=7cm,  angle=0]{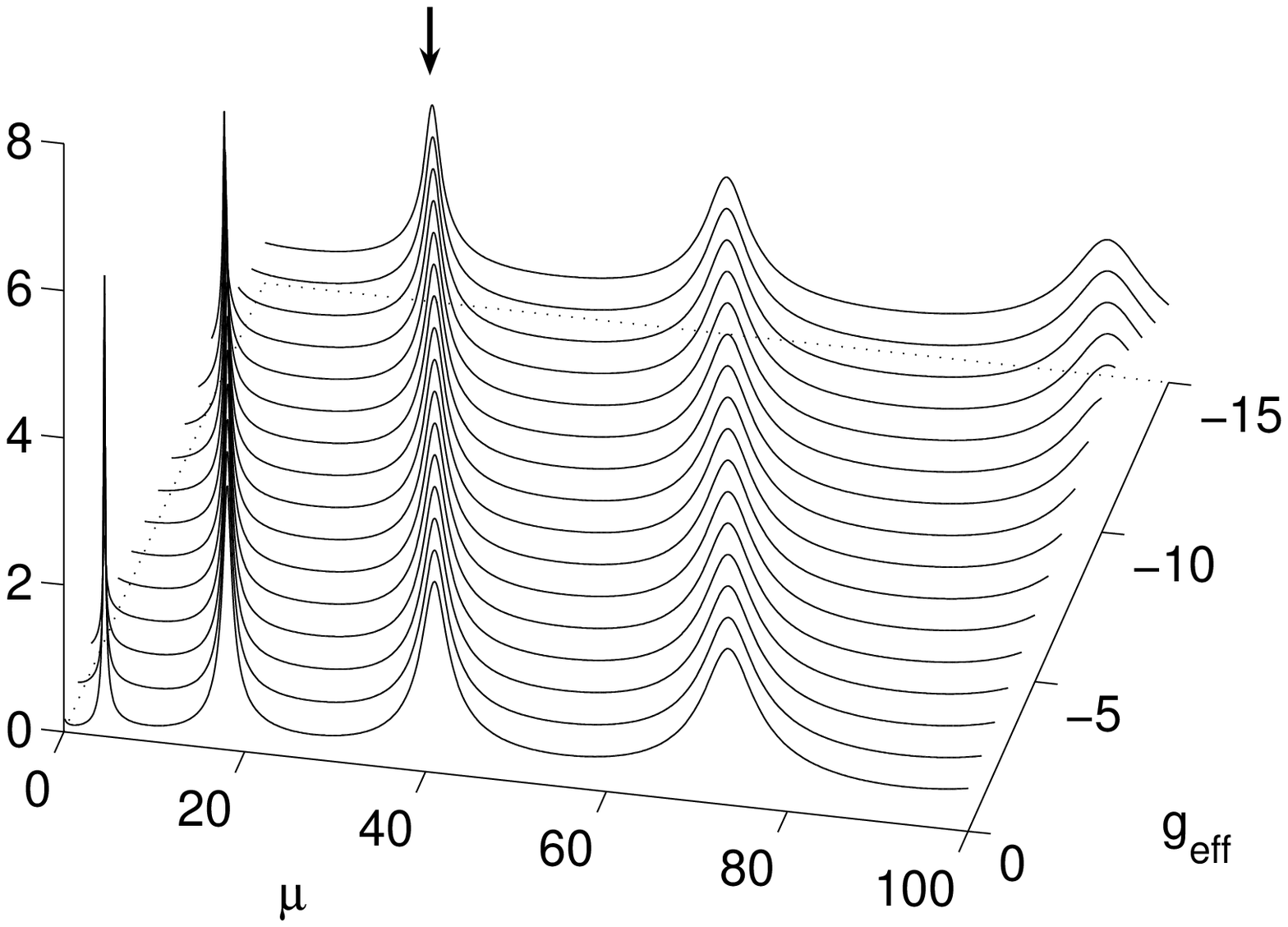}
\caption{\label{fig-attres_ratio1}
Amplitude ratio $A_l/A_r$ as a function of the chemical
potential $\mu$ for an effective nonlinearity 
$g_{\rm eff} = -5$ (left) and for different effective nonlinearities
$g_{\rm eff}$ (right).
The shift of the resonance marked by an arrow is
displayed as a function of $g_{\rm eff}$ in 
figure \ref{fig-attres-mures1}.
}
\end{figure}

In order to identify and analyze resonances of the nonlinear
Schr\"odinger equation we now calculate the amplitude ratio
$A_l/A_r$ as a function of the the chemical potential for
different values of the effective nonlinearity $g_{\rm eff}$.
The left--hand side of figure \ref{fig-attres_ratio1} shows
the amplitude ratio as a function of $\mu$ for
an effective nonlinearity $g_{\rm eff} = -5$.
As in the linear case, illustrated in figure \ref{fig_dshell_lin_resonance},
resonances can be clearly identified as maxima of the amplitude ratio
$A_l/A_r$. 
The resonances are, however, shifted to smaller values of
$\mu$, whereas the width of the resonances remains similar.

On the right--hand side of figure \ref{fig-attres_ratio1} the 
amplitude ratio $A_l/A_r(\mu)$ is plotted for different values of the effective
nonlinearity $g_{\rm eff}$. Resonances are clearly identified for
all values of $g_{\rm eff}$, but the shift of the resonance positions is
clearly visible in this illustration.
We note that the resonance heights barely change with $g_{\rm eff}$.

The observed shift of the resonances will be explained in
the following.
For convenience we rather calculate the chemical potential
where $A_l = A_r$ at the sides of each resonance, in
dependence of $g_{\rm eff}$.
These values of the chemical potential will be denoted
$\mu_n^<$ and $\mu_n^> \, $ in the following.
They are easier to calculate than the resonance positions 
$\mu_n$ because $p_l = p_r$
holds at these values, furthermore this calculation will also
reveal the influence of $g_{\rm eff}$ on the resonance width.
We note that the wavefunction on the interval $x \in [0,2a]$
is symmetric (antisymmetric) around $x=a$ for $ \mu = \mu_n^<$
($\mu = \mu_n^>$).

Using both equations (\ref{eqn-dshell-attres-amp-mu}), the chemical
potential can be written as
\be
  \mu = g A^2 \left( 1 - \frac{1}{2p} \right).
  \label{eqn-attres-mu2}
\ee
The elliptic parameter $p$ can be calculated from the relation
\be
  p K(p)^2 = \frac{|g| A^2 L^2}{16}.
\ee
Solving this relation for $p$ leads to
\be
  p = \frac{|g| A^2 L^2}{4 \pi^2} - \frac{1}{2} \left(\frac{|g| A^2 L^2}{4 \pi^2}\right)^2
  + \mathcal{O}(g^3 A^6) \, ,
  \label{eqn-attres-plarger}
\ee
and inserting this into equation (\ref{eqn-attres-mu2}), we find the desired
dependence of the chemical potential on the nonlinear interaction
\be
  \mu = \frac{2 \pi^2}{L^2} \left(1 + \frac{3 g A^2 L^2}{8 \pi^2}
  + \mathcal{O}(g^2 A^4)) \right).
  \label{eqn-attres-muA}
\ee

Formula (\ref{eqn-attres-muA}) is valid for both $\mu_n^>$ and $\mu_n^<$. Now we
insert the specific values of the period $L$ and replace $g A^2$ by the effective
nonlinearity $g_{\rm eff}$.
At $\mu_n^>$ the period of the wavefunction is $L_n^> = 2a/n$, i.e.~$\psi(a) = 0$.
Equation (\ref{eqn-geff}) for the effective nonlinearity can be easily
evaluated in lowest order in $p$, since then the elliptic function cn
equals a cosine, which yields $g_{\rm eff} \approx  g A^2 a/2$.
This finally leads to
\be
   \mu_n^> \approx \frac{n^2 \pi^2}{2 a^2} + \frac{3 g_{\rm eff}}{2 a} \, .
    \label{eqn-attres-mul-approx}
\ee

\begin{figure}[htb]
\centering
\includegraphics[width=7cm,  angle=0]{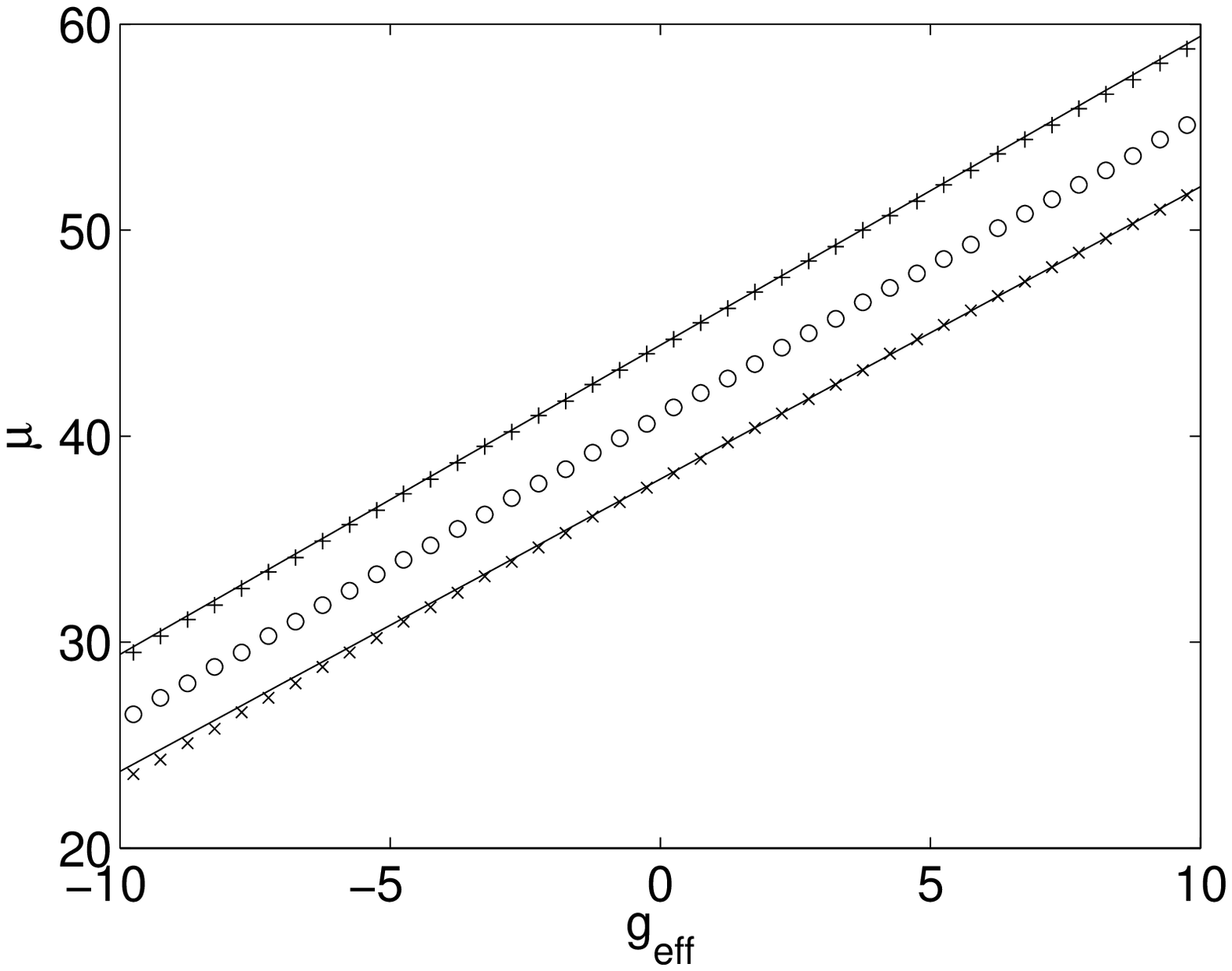}
\hspace{5mm}
\includegraphics[width=7cm,  angle=0]{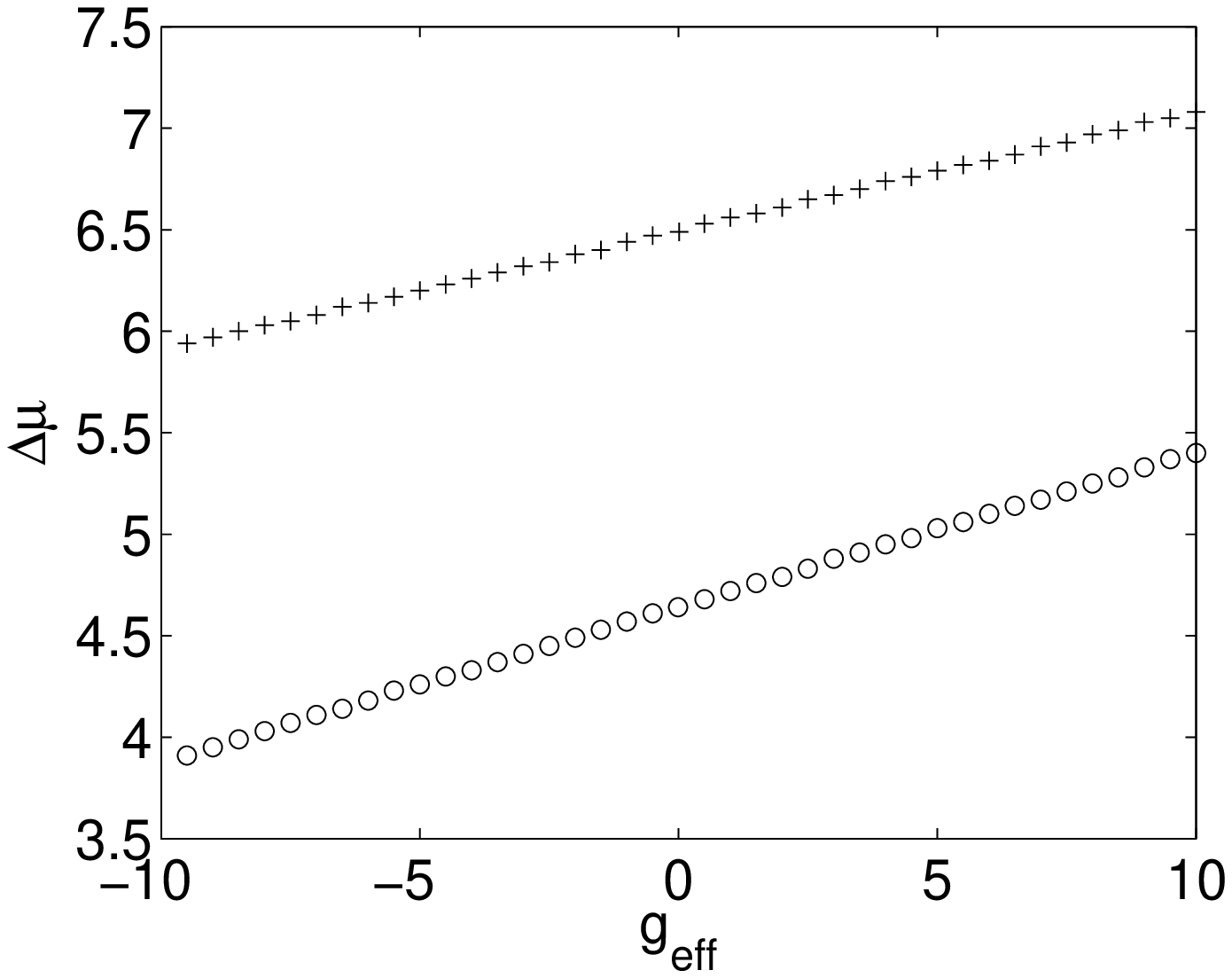}
\caption{\label{fig-attres-mures1}
Left: Dependence of the resonance position $\mu_n$ (o) and
$\mu^>_n$ (+) and $\mu^<_n$ (x) on the effective nonlinearity
$g_{\rm eff}$ for $n=3$. The solid lines are the approximations
(\ref{eqn-attres-mul-approx}) and (\ref{eqn-attres-mus-approx}).
Right: Width of the resonance, defined as $\Delta \mu = 
\mu_3^> - \mu_3^<$ (+) and as FWHM (o).}
\end{figure}

Similarly one obtains an expression for $\mu_n^<$.
In the linear case the period $L_n^<$ is given by the solution of the implicit
equation
\be
  \tan\left( \frac{2 \pi a}{L_n^<} \right) = - \frac{2 \pi}{\lambda L^<_n} .
\ee
For the example illustrated in figure \ref{fig-attres-mures1}
($a=1$, $\lambda = 10$ and $n=3$)  one has $L_3^< = 0.7215$ .
The change of $L_n^<$ with $g_{\rm eff}$ is negligible.
Again equation (\ref{eqn-geff}) for the effective nonlinearity
is readily evaluated in lowest order in $p$ and yields
\begin{eqnarray}
\fl  g_{\rm eff} &\approx& \frac{g a A^2}{2} \left( 1 - \frac{\sin(4 \pi a / L_n^<)}{4 \pi a /
    L_n^<}  \right) 
    = \frac{gaA^2}{2} \left(1 +  \frac{1}{\lambda a (1 + (2 \pi)^2/(\lambda L_n^<)^2)} \right).
\end{eqnarray}
Inserting into equation (\ref{eqn-attres-muA}), one finally arrives at
\be
   \mu_n^< \approx \frac{2 \pi^2}{(L_n^<)^2} + \frac{3 g_{\rm eff}}{2 a}
   \left( 1 - \frac{\sin(4 \pi a /
   L_n^<)}{4 \pi a / L_n^<} \right)^{-1} \, .
    \label{eqn-attres-mus-approx}
\ee

The same results are obtained in the case of a repulsive interaction
($g>0$, see below).
Thus we compare the approximations (\ref{eqn-attres-mul-approx}) and
(\ref{eqn-attres-mus-approx}) to
the numerically exact results for $g<0$ and $g>0$ together in figure
\ref{fig-attres-mures1}. We considered the resonance with $n = 3$,
that is marked with an arrow in the figures \ref{fig-attres_ratio1}
and \ref{fig-repres_ratio1}. We observe a good agreement.
Furthermore the positions $\mu_{n=3}$ of the resonances are
displayed.

From the different scaling of $\mu_n^>$ and $\mu_n^<$ we conclude
that the resonance width also changes with the effective nonlinearity.
In fact, the width increases almost linearly with $g_{\rm eff}$ and the resonances become
slightly asymmetric. The dependence of the width $\Delta \mu = \mu_3^> - \mu_3^<$
on the effective nonlinearity $g_{\rm eff}$ is illustrated
in figure \ref{fig-attres-mures1} on the right.

It should not be concealed that also bound stated can exist
in a repulsive delta--shell potential due to the attractive
self--interaction, falling of as $\sech \big( \sqrt{-2 \mu} (x-x_0) \big)$
for $x > a$.
However we will not consider these states here as we already
discussed a similar phenomenon for the single delta potential.

\subsection{Repulsive Nonlinearity}

As stated above, the real non--singular solutions of the free nonlinear
Schr\"odinger equation with a repulsive nonlinearity can be
expressed in terms of the Jacobi elliptic function sn
\cite{Carr00a, Dago00}.
Thus we make the ansatz:
\be
\psi(x) = \left\{ \begin{array}{l}
 \psi_l(x) = A_l \, \sn \bigg( 4 K(p_l)  \frac{x}{L_l}
    \bigg| p_l \bigg) \quad \mbox{for} \; x < a \\ [2mm]
 \psi_r(x) = A_r \, \sn \bigg( 4 K(p_r)  \frac{x+x_0}{L_r}
    \bigg| p_r \bigg) \quad \mbox{for} \; x > a.
  \end{array} \right.
  \label{eqn-dshell-repulsive-ansatz}
\ee
The amplitudes $A_{l,r}$ and the periods $L_{l,r}$ are now given by
\be
  A_{l,r} = \frac{4 \, \sqrt{p_{l,r}} \, K(p_{l,r})}{\sqrt{|g|} \, L_{l,r}} \quad \mbox{and} \quad
  \mu = \frac{8(p_{l,r}+1) \, K(p_{l,r})^2}{L_{l,r}^2},
  \label{eqn-dshell-repulsive-amp-mu}
\ee
where $p_{l,r} \in [0,1]$ are the elliptic parameters of
the solution on the left ($x<a$) and on the right ($x>a$)
of the delta--shell.

The boundary condition $\psi(0) = 0$ is automatically fulfilled
by the ansatz (\ref{eqn-dshell-repulsive-ansatz}).
The remaining conditions for the wavefunction and its derivative
at $x=a$ (cf. equation (\ref{eqn-delta-condition})) read:
\begin{eqnarray}
\fl {\rm I.}  \qquad  A_l \, \sn ( u_l | p_l ) &=& A_r \, \sn (u_r  | p_r ) \\[2mm]
\fl  {\rm II.} \quad 2 \lambda A_l \, \sn(u_l  | p_l )
   &=& \frac{4 A_r K_r}{L_r} \cn (u_r  | p_r ) \dn (u_r  | p_r ) 
    - \frac{4 A_l K_l}{L_l} \cn (u_l  | p_l ) \dn (u_l  | p_l ),
  \label{eqn-dshell-repres-conditions}
\end{eqnarray}
where the abbreviations $u_l = 4 K(p_l) a/L_l$ and
$u_r = 4  K(p_r) (a+x_0)/{L_r}$ have been used.

If $\abs{A_l \sn(u_l  | p_l )} \le \abs{A_r} $ the first condition
can always be fulfilled by an appropriate choice of the
''phase shift'' $x_0$.
Inserting the first condition into the second one and using
the addition theorems of the Jacobi elliptic functions one
arrives at
\begin{eqnarray}
 && \frac{2\lambda^2}{\mu} \frac{p_l}{p_l+1} \sn^2(u_l  | p_l ) +
  \frac{p_l}{(p_l+1)^{3/2}} \frac{4 \lambda}{\sqrt{2\mu}} \cn(u_l  | p_l )
  \dn (u_l  | p_l ) \sn (u_l  | p_l )  \nn \\ [2mm]
  && = \quad \frac{p_r}{(p_r+1)^2}- \frac{p_l}{(p_l+1)^2}\, .
  \label{eqn-dshell-repres-condition2}
\end{eqnarray}
The amplitude ratio $A_l/A_r$ is given by
\be
  \frac{A_l}{A_r} = \left[\frac{p_l(p_r+1)}{(p_l+1)p_r}
  \right]^{1/2}
  \label{eqn-dshell-repres-ampratio}
\ee
in terms of the elliptic parameters. A resonant enhancement of the
amplitude ratio demands that $p_l \gg p_r$.

\begin{figure}[htb]
\centering
\includegraphics[width=7cm,  angle=0]{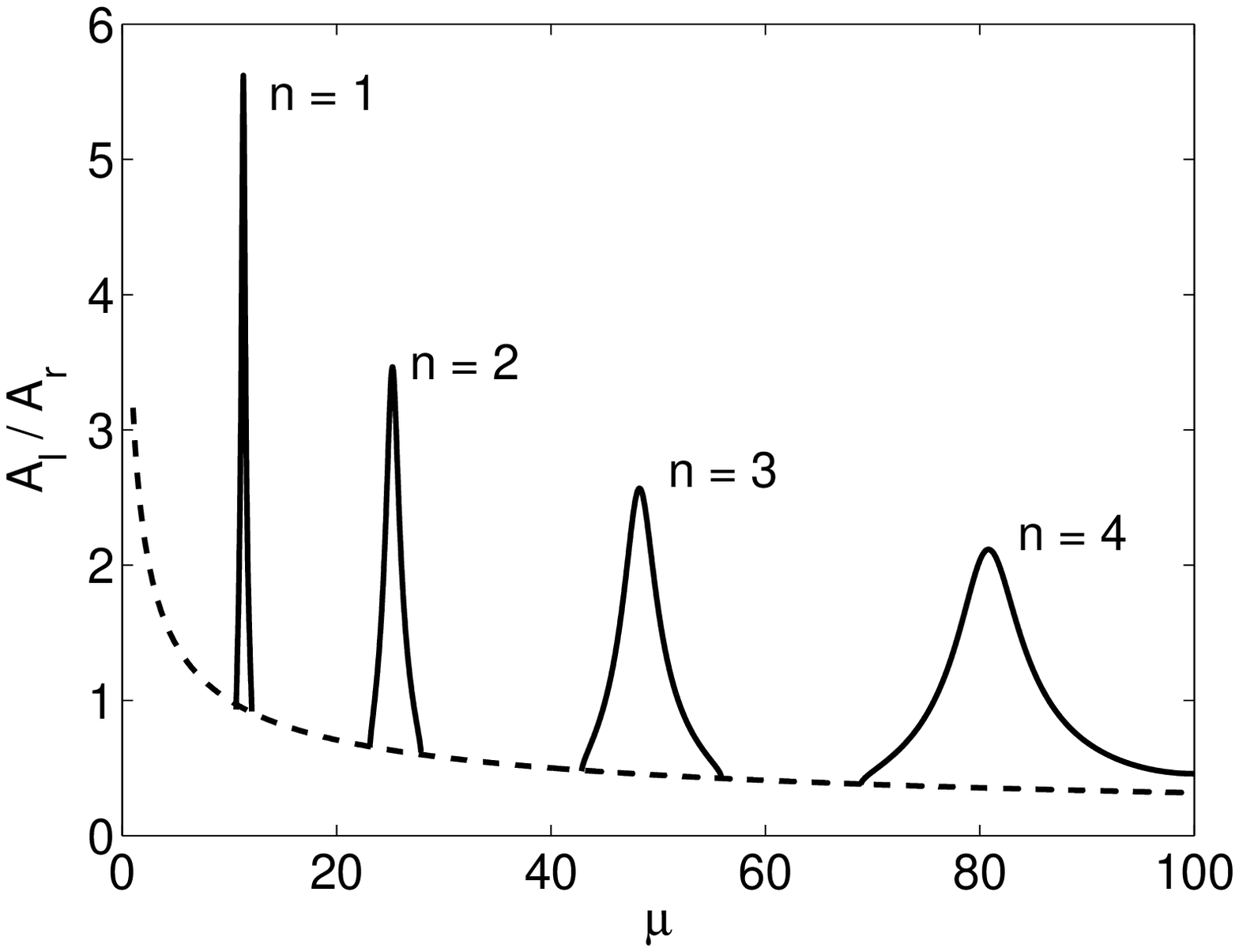}
\hspace{5mm}
\includegraphics[width=7cm,  angle=0]{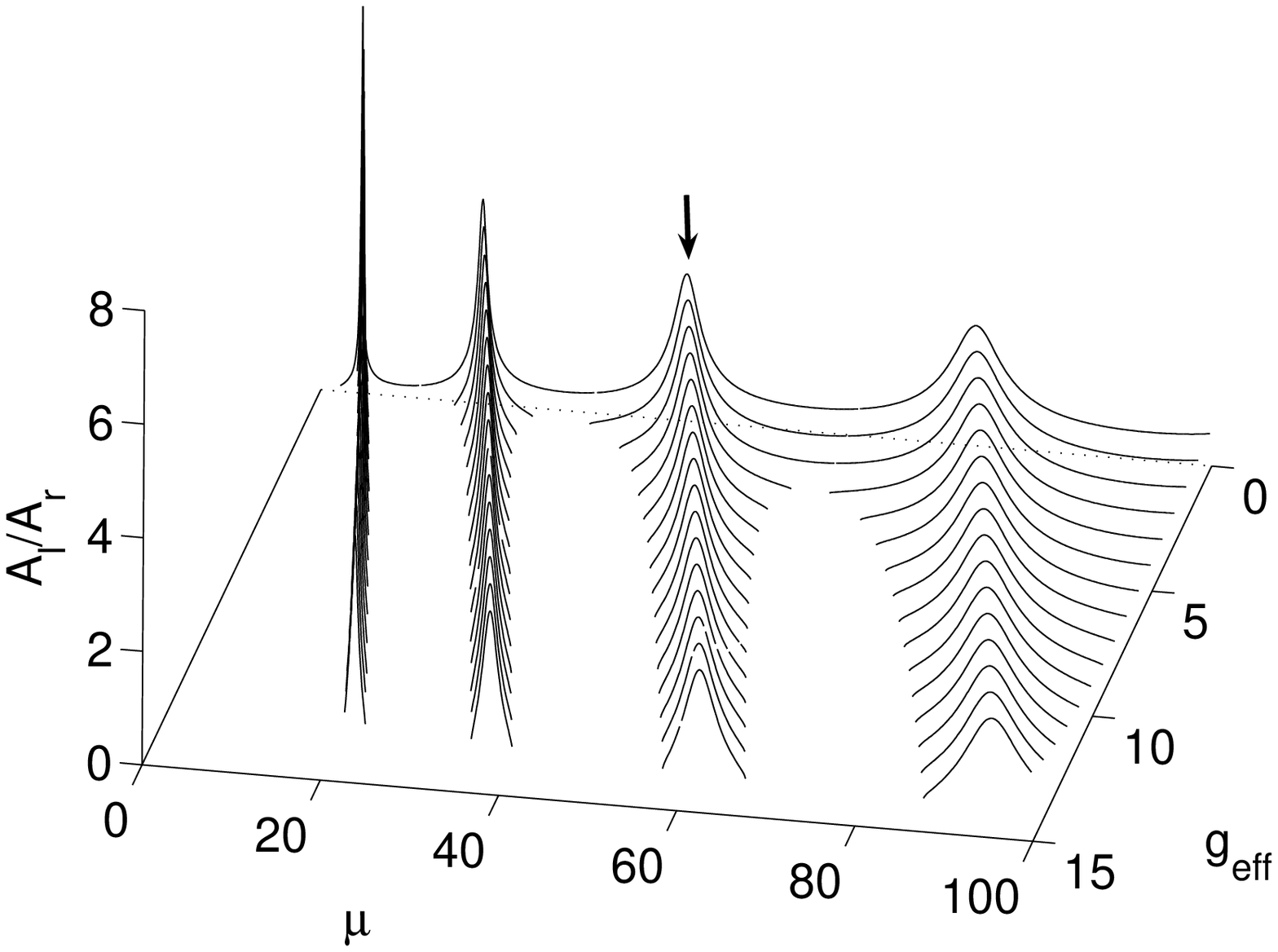}
\caption{\label{fig-repres_ratio1}
Left: Amplitude ratio as a function of the chemical
potential $\mu$ for an effective nonlinearity
$g_{\rm eff} = 5$. 
The condition (\ref{eqn-repres-condr}) for the amplitude ratio is
plotted as a dashed line.
Right: Amplitude ratio as a function of the chemical
potential $\mu$ for different effective nonlinearities.
The shift of the resonance marked with an arrow is
displayed as a function of $g_{\rm eff}$ in figure 
\ref{fig-attres-mures1}.}
\end{figure}

Again we calculated the amplitude ratio $A_l/A_r$ as a function
of the chemical potential $\mu$ for different values of the effective 
nonlinearity $g_{\rm eff}$. 
The results are illustrated in figure \ref{fig-repres_ratio1}.
The left--hand side shows the amplitude ratio $A_l/A_r$
for an effective nonlinearity $g_{\rm eff} = 5$, what should
be compared to figure \ref{fig_dshell_lin_resonance}
and figure \ref{fig-attres_ratio1}.
The first observation is that one cannot find solutions for all values
of $\mu$. In fact there exist no solutions with an amplitude ratio below 
a certain threshold. 
Resonances are still clearly identified as maxima of the amplitude ratio.
Again the resonance positions are shifted in comparison to the linear
case.

On the right--hand side the amplitude ratio is plotted for different values 
of $g_{\rm eff}$. One observes that the solutions cease to exist with an
increasing effective nonlinearity, whereas the resonances survive longest.
The resonances are shifted similarly to the case of an attractive interaction. 

Solutions with a small amplitude ratio  $A_l/A_r$
cease to exist when $g_{\rm eff}$ is increased.
In fact, condition (\ref{eqn-dshell-repres-conditions})
cannot be fulfilled any longer if the amplitude ratio $A_l/A_r$
drops  below a certain threshold.
A condition for the existence of a solution can be derived from
the equations (\ref{eqn-dshell-repres-ampratio}) and
(\ref{eqn-dshell-repulsive-amp-mu}) and yields
\be
  \left(\frac{A_l}{A_r}\right)^2 \ge  \frac{2 p_l}{p_l+1} = \frac{g A_l^2}{\mu} \, .
\ee
Inserting $g_{\rm eff} \approx g a A_l^2/2$ on the right hand side,
one is led to the approximation
\be
  \frac{A_l}{A_r} \apprge \left(\frac{2 \, g_{\rm eff}}{a \mu}\right)^{1/2} \, .
  \label{eqn-repres-condr}
\ee
As a consequence solutions apart of the resonances with small amplitude
ratios cease to exist when $g_{\rm eff}$ is increased.
This approximate condition is well confirmed by the numerical exact
results displayed in figure \ref{fig-repres_ratio1}.

The shift of the resonances is understood in the same way as in
the case of an attractive interaction. The chemical potential is
now given by
\be
  \mu = g A^2 \left( \frac{1}{2} + \frac{1}{2p} \right) ,
  \label{eqn-repres-mu2}
\ee
while equation (\ref{eqn-attres-plarger}) still holds.
Inserting into equation (\ref{eqn-repres-mu2}) and expanding
up to the linear term in $g A^2$ again leads to equation
(\ref{eqn-attres-muA}).
Thus one arrives at the same results as in the case of an repulsive
interaction, in particular at the equation (\ref{eqn-attres-mul-approx})
for $\mu_n^>$ and equation (\ref{eqn-attres-mus-approx}) for $\mu_n^<$.
The results for $n=3$ are displayed in figure \ref{fig-attres-mures1}.
The approximations agree well with the numerical exact results.
From the different scaling of $\mu_n^>$ and $\mu_n^<$ with $g_{\rm eff}$
we conclude that a repulsive nonlinearity increases the resonance width.

\section{Conclusion}

In this paper we analyzed the properties of bound, scattering and
resonance states of the nonlinear Schr\"odinger equation using two
simple model systems.

Bound, i.e.~normalizable, states were calculated and analyzed for
a single delta potential.
New features occur in the case of an attractive nonlinearity,
as states are no longer bound by an external potential but
by the internal interaction. In this case bound states can
exist despite a repulsive external potential, and they cease
to exist at a negative value of the chemical potential.
In addition we investigated the transition from bound to scattering
states.

Furthermore we discussed a repulsive delta--shell potential
as a simple model showing resonances.
Resonances can still be identified in the nonlinear case, though
the definition of a resonance becomes somewhat ambitious.
Two major effects of the nonlinearity were analyzed in detail:
Firstly, the resonance positions are shifted proportionally to the
effective nonlinearity and the resonance width increases with $g_{\rm eff}$.
Secondly, scattering states cease to exist with an increasing
repulsive nonlinearity, whereas resonances survive longest.

\section*{Acknowledgements}
Support from the Deutsche Forschungsgemeinschaft
via the Graduiertenkolleg  ``Nichtlineare Optik und Ultrakurzzeitphysik''
is gratefully acknowledged. We thank N. Moiseyev and P. Schlagheck
for stimulating discussions.


\end{document}